\documentclass{emulateapj}

\usepackage{graphicx} \usepackage{natbib} 
\shorttitle{The Resolved NIR Extragalactic Background} \shortauthors{Keenan et al.}

\begin{document}

\title{The Resolved Near-Infrared Extragalactic Background}

\author{R. C. Keenan\altaffilmark{1},
A. J. Barger\altaffilmark{1,2,3}, L. L. Cowie\altaffilmark{3},
W.-H. Wang\altaffilmark{4}}

\altaffiltext{1}{Department of Astronomy, University of Wisconsin-Madison, 475
N. Charter Street, Madison, WI 53706} 
\altaffiltext{2}{Department of Physics
and Astronomy, University of Hawaii, 2505 Correa Road, Honolulu, HI 96822}
\altaffiltext{3}{Institute for Astronomy, University of Hawaii, 2680 Woodlawn
Drive, Honolulu, HI 96822} 
\altaffiltext{4}{Institute of Astronomy and Astrophysics, Academia Sinica,
  P.O. Box 23-141, Taipei 10617, Taiwan.}

\begin{abstract}  We present a current best estimate of the integrated
  near-infrared (NIR) extragalactic background light (EBL)
attributable to resolved galaxies in $J$, $H$, and $K_s$.  Our results for
measurements of $\nu I\nu$ in units of nW~m$^{-2}$~sr$^{-1}$ are
$11.7^{+5.6}_{-2.6}$ in $J$, $11.5^{+4.5}_{-1.5}$ in $H$ and
$10.0^{+2.8}_{-0.8}$ in $K_s$.  We derive these new
limits by combining our deep wide-field NIR photometry from five widely separated fields with other studies from the literature to create a galaxy counts sample that is highly complete
and has good counting statistics out to $JHK_s \sim 27-28$.  As part of this
effort we present new ultradeep $K_s-$band galaxy counts from 22 hours of
observations with the Multi Object Infrared Camera and Spectrograph
(MOIRCS) instrument on the Subaru Telescope. We use this MOIRCS $K_s-$band
mosaic to estimate the total missing flux from sources beyond our detection
limits.   Our new limits
to the NIR EBL are in basic agreement with, but $10-20$\% higher than
previous estimates, bringing them into better agreement with estimates of the
total NIR EBL (resolved + unresolved sources)
obtained from TeV $\gamma-$ray opacity measurements and recent direct
measurements of the total NIR EBL.  We examine field to field variations in our photometry to show that the integrated light from galaxies is isotropic to within uncertainties, consistent with the expected large-scale isotropy of the EBL. Our data also allow for a robust estimate of the NIR light from Galactic stars, which we find to be $14.7\pm 2.4$ in $J$, $10.1\pm 1.9$ in $H$ and $7.6\pm 1.8$ in $K_s$ in units of nW~m$^{-2}$~sr$^{-1}$.  

\end{abstract} 
\maketitle

\section{Introduction}

The near-infrared (NIR) extragalactic background light (EBL) is the total light from resolved and
unresolved extragalactic sources in the NIR.  This represents the
integrated light from all star and galaxy formation processes over the history of the
universe that has been redshifted into the NIR.  Some fraction of the NIR EBL
can be resolved as the light from individual galaxies (Integrated Galaxy
Light, IGL), but the fractional
contribution from unresolved (and perhaps unresolvable) sources is not well
constrained.  A measurement of the total NIR EBL minus the IGL
will provide insight into the energy budget of the early universe.

While direct unresolved measurement of the NIR EBL is technically difficult due
to complex foregrounds, several authors have reported a measurement of the
total (resolved plus unresolved) NIR EBL
\citep{Haus98,Dwek98,Gorj00,Wrig01,Mats01,Mats05,Camb01,Leve07} and found it to be a factor of two
or more above the IGL obtained through source
counts \citep{Mada00,Tota01,Thom03}.
This is known as the NIR background excess (NIRBE).  The spectral energy distribution
(SED) of this measured excess, in some cases, appears very similar to that of zodiacal light,
which suggests there may be a foreground contamination issue.  Another
possible solution to this excess is a large population of undetected faint
galaxies and/or population III (PopIII) stars contributing a large fraction of
the NIR background  (see \citealt{Kash05a,Haus01} for reviews).  For example,
\citet{Mats05} found a large NIRBE with a distinct spectral
break at $\sim 1 \mu$m, which they interpreted as a possible signature of a
Lyman limit break in PopIII stellar spectra at a redshift of $\sim 10$.

Since the NIR EBL presents a source of opacity for TeV $\gamma-$rays via
pair production, the density of the background light can, in principle, be
measured via direct observation of TeV blazars.  This method relies on
assumptions of the intrinsic blazar spectrum and the SED of the EBL from
ultraviolet to the NIR, both
of which are poorly observationally constrained.  Nevertheless, TeV $\gamma-$rays provide an
independent estimate of the NIR EBL that can be used to help determine what
fraction of the background light can be attributed to resolved sources, and
how much may arise from faint and possibly exotic sources in the early
universe.  

  \citet{Dwek05a} explored the possibility that the
NIRBE is of extragalactic origin by looking for the absorption imprint of the
$\sim 1 \mu$m  spectral break on
the $\gamma-$ray SEDs of several TeV blazars.  They concluded that the
apparent NIR excess is likely not of extragalactic origin because its
signature is not detected in the blazar SEDs.  Recent observations of TeV
blazars with the High Energy Stereoscopic System (HESS) have placed limits on
the NIR EBL that allow for the possibility of  little or no excess above
integrated galaxy counts \citep{Ahar06}.  

\citet{Mazi07} used 13 TeV blazars and a grid of
NIR background SEDs to further constrain the background
light. They derived limits on the NIR EBL that are roughly in agreement with
those of \citet{Ahar06}.  However, systematic uncertainties in these
$\gamma-$ray opacity measurements due to uncertainties in the blazar SED and
that of the EBL result in large error bars that could still allow for a
significant NIRBE (of the same order as the IGL) to be present, even if their
assumptions about the intrinsic blazar SED and that of the EBL are assumed to
be reasonable.
 
 As such, the goal of this paper is to derive a firm lower limit to the NIR
 IGL so as to facilitate a robust comparison with future new results from
 direct and indirect measurements of the NIR EBL.  We present the best current
 estimate of the NIR IGL by combining our deep wide-field NIR data set with a
 compilation of galaxy counts from the literature and ultradeep $K_s-$band
 galaxy counts from a subfield of the Subaru MOIRCS image presented in
 \citet{Barg08} and \citet{Wang09}.   

The structure of this paper is as follows:  We describe our NIR dataset
briefly in Section~\ref{nird}.  We calculate the contribution of resolved
galaxies to the NIR EBL in Section~\ref{nir_resolved}.  We compare with
previous measurements of the NIR EBL and IGL and discuss the implications of
our results in Section~\ref{comparison}. We summarize in
Section~\ref{summary}.  Unless otherwise noted, all magnitudes given in this
paper are in the AB magnitude system ($m_{\rm{AB}} = -2.5\log f_{\nu}-48.60$ with
$f_{\nu}$ in units of $\rm{ergs} ~\rm{cm}^{-2}~\rm{s}^{-1}~\rm{Hz}^{-1}$).

\section{The NIR Data}
\label{nird}
We presented the observations, data reduction, star-galaxy separation, and
bright galaxy counts for our deep wide-field NIR dataset in \citet{Keen10}(K10
hereafter).
The data consist of five widely separated fields at high galactic latitudes,
including one centered on an Abell cluster (Abell 370). The survey covers
approximately 3 deg$^2$ reaching a 5~$\sigma$ limiting magnitude of $JHK_s
\sim 22-23$ over $\sim 2.75$ deg$^2$ with another $\sim 0.25$ deg$^2$ to
$JHK_s \sim 24$, making it one of the deepest wide-field surveys to date in the
NIR. The fields covered in this survey are described in Section~\ref{widefield}.

In this paper we supplement the above data with those from
an ultradeep $K_s-$band mosaic composed of imaging done by our group and
Japanese groups led by various investigators using the Multi Object InfraRed
Camera and Spectrograph (MOIRCS) instrument on the 8.2~m Subaru Telescope. The
MOIRCS imaging is described in Section~\ref{moircs}. 

\subsection{Wide-Field Imaging and Galaxy Counts}
\label{widefield}

Our first two fields are centered on the \emph{Chandra} Large Area
Synoptic X-ray Survey (CLASXS; \citealt{Yang04,Stef04}) and the \emph{Chandra} Lockman Area North Survey
(CLANS; \citealt{Trou08,Trou09}). Each of these fields cover $\sim 1$~deg$^2$ in
$JHK_s$.  These fields  are located in the Lockman Hole region of extremely
low Galactic HI column density \citep{Lock86}.  Our third field covers a
0.25~deg$^2$ area centered on the \emph{Chandra} Deep Field North (CDF-N, \citealt{Bran01, Alex03}).  The CDF-N contains the Great Observatories Origins
Deep Survey North (GOODS-N; 145 arcmin$^2$ \emph{HST}
Advanced Camera for Surveys observation, \citealt{Giav04}).  Our
fourth field is the Abell 370 (A370) cluster and surrounding area ($\sim 0.5$
deg$^2$).  A370 is a cluster of richness 0 at a redshift of $z=0.37$.  Our
fifth field is a $\sim 0.2$ deg$^2$ area  centered on the ``Small-Survey-Area
13''(SSA13) from the Hawaii Deep Fields described in \citet{Lill91}. 

In K10 we used the bright NIR galaxy counts from our survey in
combination with counts we generated from The Two Micron All Sky Survey
(2MASS,~\citealt{Skru06}), The UKIRT Infrared Deep Sky Survey (UKIDSS,
\citealt{Lawr07}) and other selections from the literature to explore
local large scale structure via the slope of the galaxy counts curve as a
function of position on the sky.  Here we extend the counts from
K10 to fainter magnitudes using the methods described below.

We generated average galaxy counts for our ensemble of NIR images by the
methods described in K10.    
We corrected the counts for completeness
through a simulation in which relatively bright ($JHK_s = 19$)  galaxies were extracted from each image and
then scattered at random to other parts of the image to test how well they were
recovered as a function of apparent magnitude.  We faded the test galaxies
in 0.25 magnitude steps until the recovered fraction fell below 25\%.  We only
included counts in the final average where completeness was better than
50\%.

\begin{figure}
\begin{center}
\includegraphics[width=90mm]{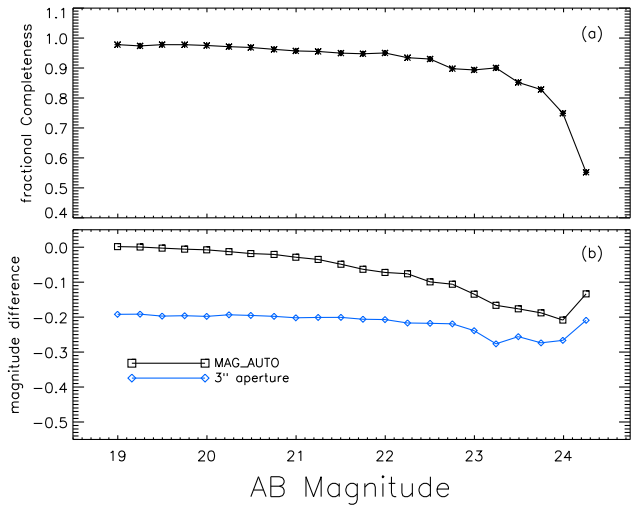}
\caption{\label{complete} Example results from our completeness
  simulations (this case is for the CDF-N $K_s-$band, though results are similar
  across all fields).  (a) Fractional completeness as a function of apparent
  magnitude.  Asterisks and solid
  black line show fractional completeness ($N_{recovered}/N_{scattered}$) as a
  function of apparent magnitude (b) Magnitude difference as a function of
  apparent magnitude comparing the MAG\_AUTO and 3$\arcsec$ apertures and
  showing total magnitude minus recovered magnitude for the two different apertures.  Black squares
  show results for the MAG\_AUTO aperture and blue diamonds for a 3$\arcsec$ aperture.  The MAG\_AUTO aperture does a good job at recovering total
  magnitude for objects brighter than 20$^{th}$ magnitude, and the 3$\arcsec$
  aperture shows a roughly constant offset ($\sim -0.2$) from total
  magnitude.}
\end{center}
\end{figure}

In our completeness simulations, given that we know the total magnitude of
objects we are scattering throughout the image, we are able to compare the
recovered magnitudes in various apertures to total magnitudes and thereby
derive a magnitude correction factor for faint objects.  Figure~\ref{complete}
shows example results from this exercise.  In Figure~\ref{complete}a we show the fractional
completeness ($N_{recovered}/N_{scattered}$) as a function of
apparent magnitude for the CDF-N $K_s-$band. In all fields we reliably
recover, on average, relatively bright objects ($JHK_s < 20$) at the correct magnitude using the
SExtractor MAG\_AUTO, which fits a Kron-like elliptical aperture to
each object \citep{Kron80}. We find that the 3$\arcsec$
aperture shows a roughly constant offset ($\sim -0.2$) from total
magnitude.  Based on these results, we measure total magnitudes in the
MAG\_AUTO aperture for objects brighter than 20$^{th}$ magnitude and switch
to corrected 3 arcsecond aperture magnitudes for objects fainter than
20$^{th}$ magnitude using the measured difference at 20$^{th}$ magnitude.
This magnitude correction only affects our calculated IGL at the
$0.1$ nW m$^{-2}$ sr$^{-1}$ level, which is roughly a factor of 10 below the
uncertainty in the IGL due to the uncertainty in the counts themselves.

\subsection{MOIRCS Imaging and Galaxy Counts}
\label{moircs}
The Subaru MOIRCS $K_s-$band image is centered on the GOODS-N field \citep{Giav04}.  Part of the Japanese
data were published in \citet{Kaji06}.  Here we include all data taken
between 2005 and 2008.  The data reduction for the MOIRCS mosaic is described in
\citet{Barg08} and \citet{Wang09} and is essentially the same as that used in
K10.  

$K_s-$band galaxy counts from the MOIRCS mosaic presented
here are taken from a subfield of $\sim 15$ arcmin$^2$ with an integration
time of $\sim 22$ hr yielding a $5~\sigma$ limit of $\sim 26.5$.  This MOIRCS
image is, therefore, approximately as deep as the Subaru Deep Field (SDF,
\citealt{Maih01}) and Subaru Super Deep Field (SSDF, \citealt{Min05}) but more
than an order of magnitude larger in area on the sky than either survey.    

We generated a catalog for the MOIRCS field using SExtractor software version 2.4.4 \citep{BA96} for source identification and
photometry.   We used the MAG\_AUTO aperture and set the minimum number of contiguous pixels to be considered
a detection (SExtractor parameter DETECT\_MINAREA) to 2 in order to push the
counts on the faint end.  We did not remove stars from the MOIRCS
counts because we begin counting galaxies at 23$^{rd}$ magnitude where stars
identifiable through color selection do not contribute significantly to galaxy
counts ($\sim 1$\% for $JHK_s = 23$ from K10 star counts compared
with average galaxy counts from this paper).  Cool dwarf stars unidentifiable
through color selection may contaminate the counts on a slightly higher level,
but their space density is unknown and we have no means of separating them
from galaxies.  

For the Subaru MOIRCS $K_s-$band data, we do not correct for completeness or
total magnitudes.  Rather, we calculate the total missing flux
in the image using a method similar to that employed by \citet{Thom07a}.  We
discuss this in more detail in Section~\ref{missingflux}. We adopted this
method because it allows us to not only account for the flux in objects missed
near our detection limit (completeness correction), but also for the flux missed
in objects or parts of objects below our detection limit.  In $J-$band and
$H-$band we adopt the missing flux estimates of \citet{Thom07a}.

\begin{figure*}
\begin{center}
\includegraphics[width=180mm]{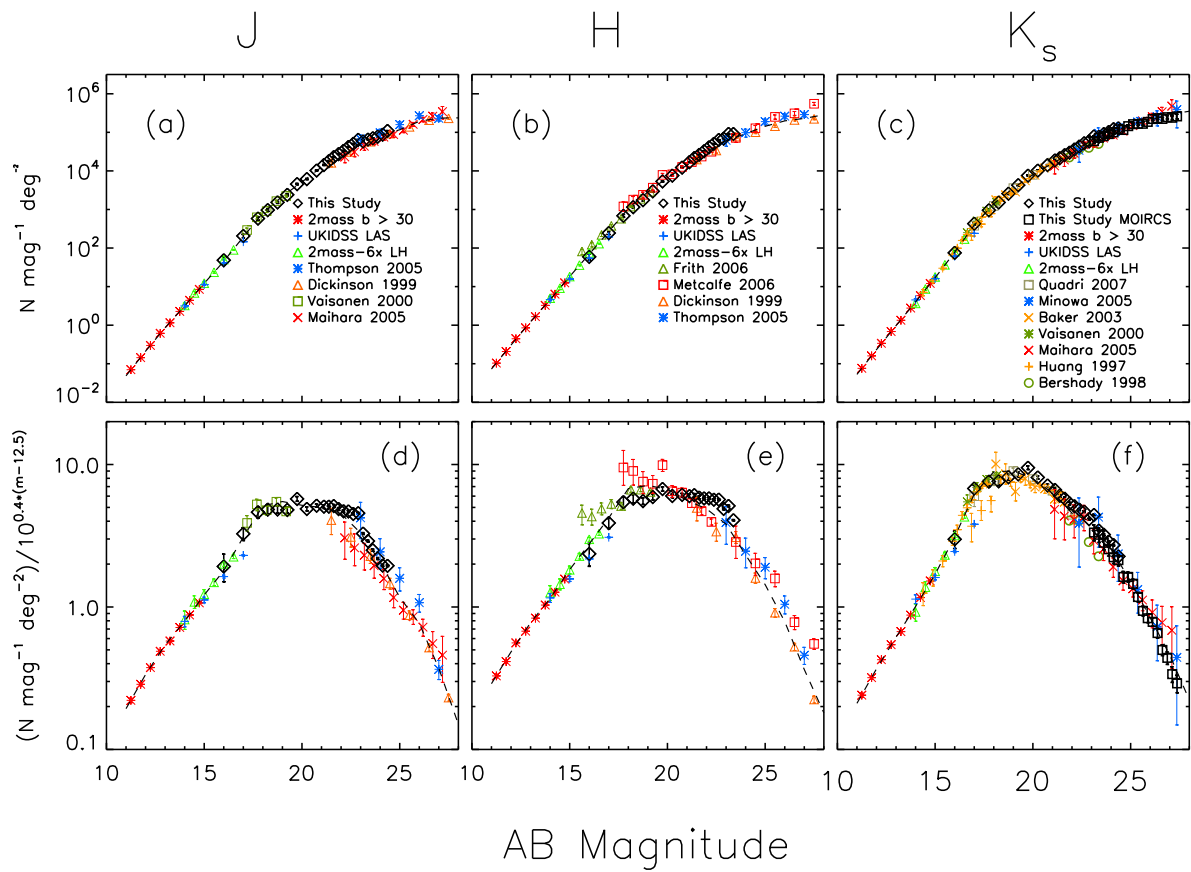}
\caption{\label{mycounts}(a-c)   Galaxy counts as a function of apparent
  magnitude.  Average galaxy counts from our deep,
  wide-field data on 5 fields (this study) are denoted by black diamonds.  The $K_s-$band counts from our Subaru MOIRCS data
  (this study MOIRCS) are denoted by black squares.  Error bars for this work are approximately the
  size of the plot symbols.  The counts determined by K10 from the
  2MASS field with Galactic latitudes of $|b|>30$ are denoted by red
  asterisks. The counts determined by K10 from the the 2MASS-6x
  Lockman Hole survey are denoted by green triangles.  The counts determined
  by K10 from three subfields of the UKIDSS LAS are denoted by blue
  plus symbols. Other data points are
  taken from the studies listed in the plots.  The dashed curve shows our
  error-weighted least squares running average (described in
  Section~\ref{nir_resolved}) from which we calculate the NIR IGL.  (d-f) The
  same data as in (a-c) but divided by an arbitrarily normalized Euclidean
  model with slope $\alpha = 0.4$. }
\end{center}
\end{figure*} 

\section{NIR Background Due to Resolved Galaxies}
\label{nir_resolved}
\begin{figure}
\begin{center}
\includegraphics[width=90mm]{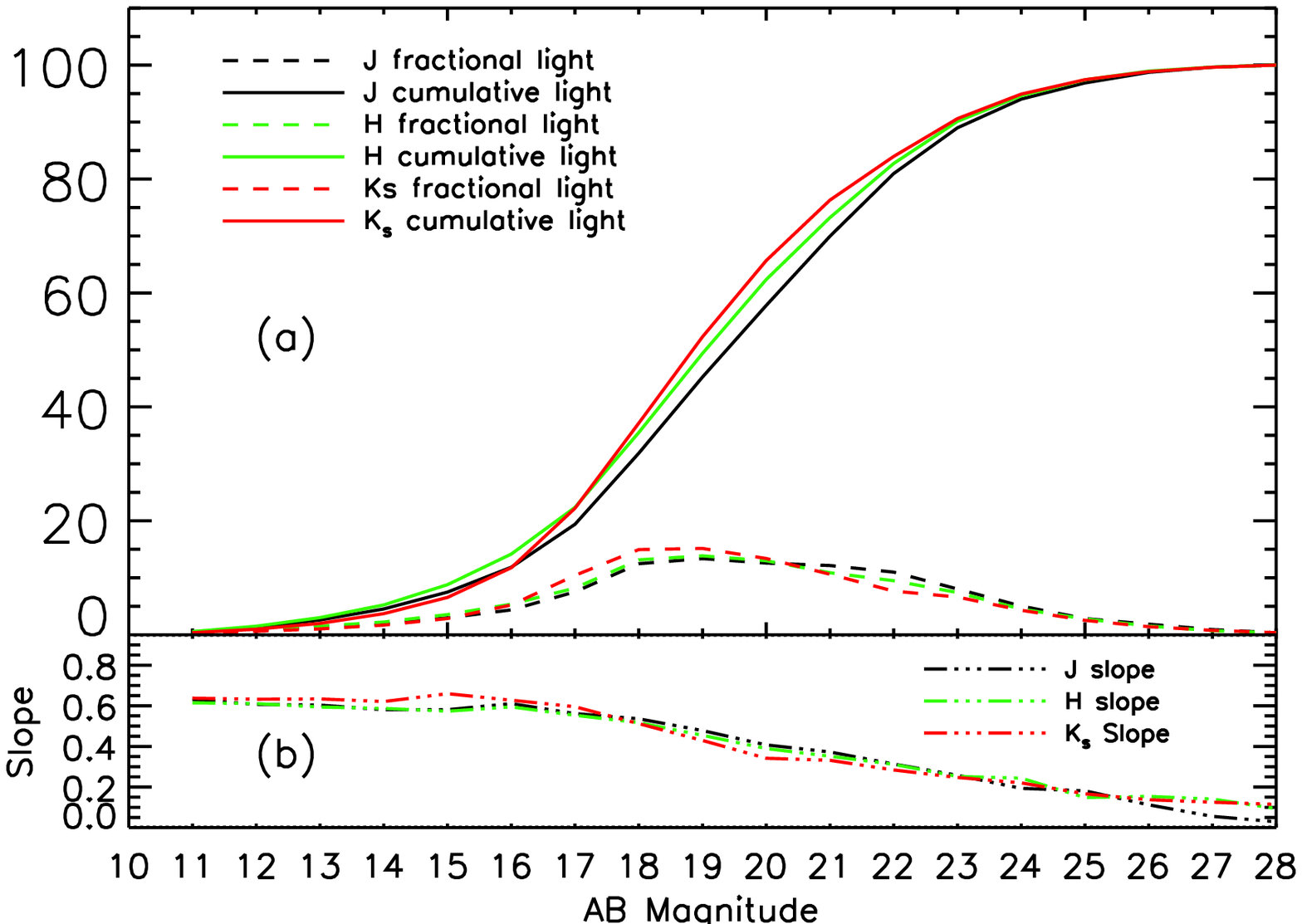}
\caption{\label{nirebl_calc}(a) Percent contribution to the IGL as a function
  of apparent magnitude.  Solid curves show the cumulative contribution
  from galaxies to the NIR IGL as a function of apparent magnitude.  Dashed
  curves show the fractional contribution per each magnitude bin. This
  demonstrates that the vast majority of the IGL ($\sim90$\%) arises from
  galaxies in the apparent magnitude range $\sim 15 < JHK_s < 24$ where this
  study is highly complete and has good counting statistics.  (b)  Slope of
  the galaxy counts curve as a function of apparent magnitude.
  Dash-dotted curves show the measured slope of the galaxy counts curve as a
  function of apparent magnitude.  When the slope drops below $\alpha = 0.4$
  the total light from galaxies becomes convergent.  }
\end{center}
\end{figure} 
In Figure~\ref{mycounts} we show our completeness-corrected and averaged
galaxy counts with those drawn from the literature.  However, before combining
all of our data in this way, we first investigated field to field variations for the IGL for
our five fields over the magnitude range $14.5 < JHK_s < 22.5$ where all five
are highly complete and have good counting statistics.  We found the
IGL over this magnitude range to be consistent across the four non-cluster
fields (CLASXS, CLANS, CDF-N and SSA13 $\sim 7-8$
nW~m$^{-2}$~sr$^{-1}$) with a $1~\sigma$
dispersion of $\pm 0.5$
nW~m$^{-2}$~sr$^{-1}$.  As such, we find that the IGL is consistent with large-scale istropy, an expected signature of the EBL (see \citealt{Kash05a,Haus01} for reviews).

In the A370 cluster field, our IGL measurements were
$\sim 2-3$ nW~m$^{-2}$~sr$^{-1}$ higher in all bands with the peak
contribution to the excess light arising from galaxies at $JHK_s \sim 17$,
consistent with an excess of $L_*$ galaxies at a redshift of $\sim 0.4$ as in
A370.  We include the cluster field in our average counts for this study
because in a survey of a few square degrees such as this, roughly one massive
cluster will be present.

In Figure~\ref{mycounts} we use black
diamonds to denote the counts for our five wide fields and black squares to
denote counts for the MOIRCS image. The counts determined by K10
from the 2MASS field with Galactic latitudes of $|b|>30$ are denoted by red
asterisks. The counts determined by K10 from the the 2MASS-6x
Lockman Hole survey are denoted by green triangles.  The counts determined by
K10 from three subfields of the UKIDSS LAS are denoted by blue plus
symbols.   The 
three subfields that make up this section of the LAS are described in detail
in K10.  Other data points are taken from the publications listed
in the plots.  The dashed curves represent the error-weighted least squares
running average from which we calcuate the IGL.  We describe the methods
we used to calculate this average in Section~\ref{nir_resolved}.  

\begin{deluxetable*}{ccccccc}
\tablecaption{\label{avgcntstab}Average Galaxy Counts Used to Calculate the NIR EBL}
\tablehead{  & \multicolumn{2}{c}{~~~$J$~~~} &\multicolumn{2}{c}{~~~$H$~~~} & \multicolumn{2}{c}{~~~$K_s$~~~}}  
\startdata
Mag(AB)\tablenotemark{a}  & Log$_{10}$($N$)\tablenotemark{b} & Log$_{10}$($\Delta N$)\tablenotemark{c} & Log$_{10}$($N$)  & Log$_{10}$($\Delta N$) &  Log$_{10}$($N$)  & Log$_{10}$($\Delta N$)  \\
\hline
11&-1.31&-2.33&-1.13&-2.67&-1.27&-2.76  \\
12&-0.67&-2.10&-0.51&-2.02&-0.63&-1.94  \\
13&-0.06&-1.27&0.089&-1.60&-0.01&-1.64  \\
14&0.500&-1.30&0.660&-0.25&0.603&-0.92  \\
15&1.084&0.121&1.262&-0.32&1.229&0.553  \\
16&1.654&0.726&1.842&1.004&1.891&0.379  \\
17&2.290&1.517&2.425&1.519&2.592&2.013  \\
18&2.910&2.365&3.030&1.786&3.149&1.851  \\
19&3.340&2.665&3.453&2.644&3.554&2.228  \\
20&3.713&3.117&3.825&2.736&3.901&2.479  \\
21&4.098&3.495&4.148&3.368&4.200&2.701  \\
22&4.455&3.811&4.488&3.476&4.459&2.946  \\
23&4.719&4.162&4.781&4.059&4.795&3.669  \\
24&4.916&4.391&4.965&4.170&5.008&3.931  \\
25&5.067&4.242&5.156&4.732&5.178&4.653  \\
26&5.285&4.439&5.297&4.647&5.324&4.618  \\
27&5.370&4.297&5.361&5.055&5.462&5.227  \\
28&5.376&4.717&5.456&4.755&5.547&4.983  \\
\enddata
\tablenotetext{a}{AB Magnitude.}
\tablenotetext{b}{$N$ is the surface density of galaxies in units of $\rm{mag}^{-1}\rm{deg}^{-2}$.}
\tablenotetext{c}{$\Delta N$ is the $1~\sigma$ error estimate for N (in
  the same units) derived from
  the fits described in Section~\ref{nir_resolved} to all the data shown in Figure~\ref{mycounts}.}
\end{deluxetable*}

Figures~\ref{mycounts}(d-f) display the same data as in
panels (a-c) after dividing by an arbitrarily normalized
Euclidean model (of $\alpha = 0.4$ in the form $N(m) = A \times$ 10$^{\alpha m}$, and $A$ is a
constant) to expand the ordinate and demonstrate where resolved galaxies
contribute the most to the IGL.  A flat line in (d-f) would imply an equal
contribution to the IGL at all magnitudes.  The areas of positive slope
show where galaxies contribute a larger fraction to the IGL as one moves
toward fainter magnitude.  The steep negative slope beyond $JHK_s > 23-24$
demonstrates the diminishing contribution of resolved galaxies to
the IGL at the faintest magnitudes.

To calculate the IGL from  all the counts data displayed in
Figure~\ref{mycounts}, we used an error-weighted least-squares fit of the
the slope and normalization parameters ``A'' and ``$\alpha$'' in the function
$N(m) = A\times 10^{\alpha m}$. We fit
the data over a range of $\pm 2$ magnitudes above and below each magnitude bin to get two estimates for the
counts in that bin.  In other words, at a given
apparent magnitude ``$m$'', we fit the range $m-2$ to $m$ and $m$ to $m+2$
separately.  Each fit yielded an estimate of the counts at apparent
magnitude $m$.  We then recorded the average of these two values as the best
estimate for counts in that bin.  Each of the two fits also yielded a $1~\sigma$
confidence interval for the counts in each bin.  Given these fitted values and
confidence intervals we found the maximum and minimum allowable counts over
the $1~\sigma$ range and recorded $\frac{maximum-minimum}{2}$ as the final
$1~\sigma$ error associated with the average value for the counts in that bin.
We then integrated these fitted values for the galaxy counts to
calculate the IGL and its associated $1~\sigma$ confidence interval for the
magnitude range $11 < JHK_s < 28$. 
Our fitted values for the galaxy counts  with the confidence intervals as
described above are given in tabular form in Table~\ref{avgcntstab}.

The results of this calculation are shown in Figure~\ref{nirebl_calc}.  In
Figure~\ref{nirebl_calc}a, the solid curves represent the cumulative
contribution of galaxies to the IGL as a function of apparent magnitude,
and the dashed curves show the percent contribution in each one-magnitude bin.
Figure~\ref{nirebl_calc}b shows the calculated slope of the galaxy counts
curve as a function of apparent magnitude.  Due to simple geometrical effects,
when the slope drops below $\alpha = 0.4$ the total light from galaxies
begins to converge.  

Figure~\ref{nirebl_calc}b shows that the counts slope for all three bands
begins near the expected Euclidean value of $\alpha = 0.6$ at bright
magnitudes and then drops steadily beyond $JHK_s \sim 17$.  Due to these
trends in slope, the vast majority ($\sim 90\%$) of the resolved IGL
arises from galaxies in the range $\sim 15 < JHK_s < 24$.  Over this entire
range our study is highly complete and has good counting statistics.

Furthermore, from Figure~\ref{nirebl_calc} it can be seen that for sources
fainter than $JHK_s = 28$ to make any appreciable contribution to the IGL
the counts curve would have to rise dramatically over several magnitudes
beyond the limits of current surveys.  Such a rise would quickly result in
several (or more) galaxies per square arcsecond, rendering any counting exercise
impossible due to confusion.  Faint galaxies certainly exist beyond the limits
of the deepest NIR surveys, because at high redshifts the faintest apparent
magnitudes observed are only probing a few magnitudes fainter than $L_*$ down
the luminosity function.  It is unknown whether there is a steep upturn in the
luminosity function toward faint magnitudes, but if such an upturn exists and
faint galaxies contribute significantly to the NIR EBL, they would need to be
so numerous as to be unresolvable.
As such, it may be safe to say that the
resolvable portion of the NIR EBL (the IGL) has, for the most part, been measured and
that the most important contribution to the resolved portion comes from
galaxies in the magnitude range $\sim 15 < JHK_s < 24$, for which the deep
wide-field data presented here are optimized.   

\subsection{$K_s-$band Faint Counts Slope and Missing Flux}
\label{missingflux}
We use our ultradeep MOIRCS data to test for a possible
steepening of the counts slope at faint magnitudes in the $K_s-$band, as
appeared possible from the SDF \citep{Maih01} and SSDF
\citep{Min05} $K_s-$band counts.  We see from Figure~\ref{mycounts}c, that the
slope of the counts in the MOIRCS field continues to become flatter out
to the faintest magnitudes.  We detect no upturn in the slope for very faint galaxies.  As
mentioned above, the MOIRCS counts have not had a completeness correction
applied.   Such a correction would steepen the curve slightly at the faintest
magnitudes, but this effect would not be strong enough to alter significantly
the trend in slope. 

\begin{figure}
\begin{center}
\includegraphics[width=90mm]{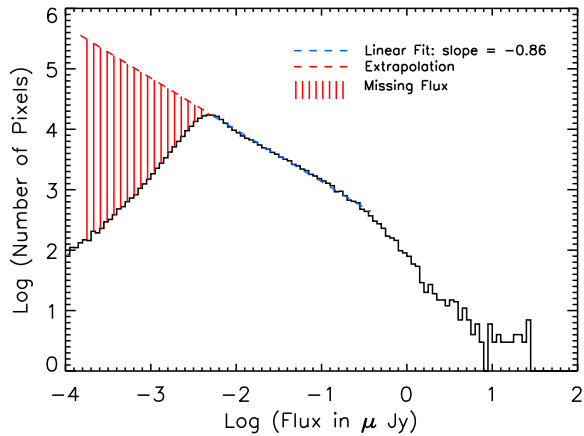}
\caption{\label{missingkflux} Histogram showing Log$_{10}$ of flux in $\mu$Jy
  vs. Log$_{10}$ of the number of pixels at that flux for all the pixels
  associated with galaxies in the Subaru MOIRCS $K_s-$band image.  The blue dashed line
  shows a linear fit from $\sim 0.005 - 0.4~\mu$Jy, which includes $\sim~60\%$
  of all pixels associated with galaxies and $\sim~99\%$ of such
  pixels containing fluxes higher than the peak of the histogram.  The slope
  of this line is -0.86.  We extrapolate the linear fit toward fainter fluxes to
the point where all the pixels in the image are accounted for (red dashed
line).The red hashed area shows the missing flux component corresponding
to 1.9 nW~m$^{-2}$~sr$^{-1}$ in the $K_s-$band.}
\end{center}
\end{figure}

\citet{Thom07a} estimated the missing flux component in the $J$ and $H-$bands
from the faint outer parts of galaxies and from galaxies below their detection
limits using a histogram of flux in all pixels associated with detected
objects.  We use a similar method to estimate the flux missed in the
$K_s-$band.  In Figure~\ref{missingkflux} we show a histogram of number
of pixels versus flux for all pixels associated with detected galaxies (object pixels) in our MOIRCS
$K_s-$band mosaic.  Noting the linear trend for fluxes $\sim
0.005 - 0.4~\mu$Jy, we fit a line to the data over this range (blue dashed
line). The portion of the histogram used in the fit represents 60\% of all
object pixels in the image and 99\% of object pixels with fluxes greater than
the turnover in the histogram at $\sim 0.005~\mu$Jy.  
The slope of the linear fit is -0.86.  \citet{Thom07a} find a slightly steeper
value of -1 by simply estimating the slope by eye.  Assuming that the true flux distribution for
faint pixels ($< 0.005~\mu$Jy) continues along the same trend, we extrapolate
the linear fit to approximate the shape of the histogram when all pixels in
the image are accounted for. Using this method, we calculate an estimate for
flux missed in the faint outer parts of galaxies and in galaxies that are
below our detection limits.  We find the missing flux component to be $\sim
22\%$ (1.9 nW~m$^{-2}$~sr$^{-1}$) of the total $K_s-$band light from resolved
galaxies. 
\subsection{The Stellar Component of NIR Light}
The integrated intensity from stars in the NIR (at Earth) is comparable to
that of resolved galaxies (as determined by models, such as those of
\citealt{Leve07}, or by integrated star
counts, as discussed below). Thus, in terms of measuring the IGL, it is
crucial to be able to separate stars from galaxies accurately.  In K10 we
demonstrate that  we are robustly separating stars from galaxies in our
analysis.  Thus, we are able to derive the stellar contribution to total NIR
light with star counts from $15 < JHK_s < 24$, while groups making unresolved
measurements of the total NIR EBL must assume stellar distribution models to
remove the NIR light due to stars from their measurements.  The integrated
starlight from our star counts is 14.7 $\pm 2.4$ nW m$^{-2}$ sr$^{-1}$ in
$J-$band,  10.1 $\pm 1.9$ nW m$^{-2}$ sr$^{-1}$ in
$H-$band, and 7.6 $\pm 1.8$ nW m$^{-2}$ sr$^{-1}$ in
$K_s-$band.  

Four of our five fields are at Galactic latitudes of $b \sim 50$, and one is at
$b \sim 70$ (we average over all five in the numbers quoted above).  Looking at the modeled starlight as a function of Galactic
latitude in \citet{Leve07}, we find the average NIR light due to stars in
their models at these latitudes is $\sim 7$ nW m$^{-2}$ sr$^{-1}$ in
$K-$band and $\sim 16$ nW m$^{-2}$ sr$^{-1}$ in
$J-$band, in good agreement with our measured values. Thus, it
appears that recent total NIR EBL measurements are removing the stellar
component correctly.  However, as pointed out in K10, cool dwarf
stars may contaminate the galaxy counts fainter than $JHK \sim 19$, suggesting
that perhaps both IGL measurements and direct total NIR EBL
measurements retain some level of stellar contamination, but this would not
affect any discrepancy between the two.  TeV $\gamma-$ray observations are
concerned with the NIR background along the entire path length to the blazars
and are unaffected by Galactic NIR starlight due to its local origin.

\section{Comparison With Previous Results}
\label{comparison}
\begin{deluxetable*}{llll}
\tablewidth{0pt}
\tabletypesize{\scriptsize}
\tablecaption{\label{nirebl_tab}NIR Integrated Galaxy Light ($\nu I_{\nu};~\rm{nW}~\rm{m}^{-2}~\rm{sr}^{-1}$)}
\tablehead{ ~~~~~& $J$\tablenotemark{a} & $H$ & $K_s$\tablenotemark{b}} 
\startdata
\scriptsize{This work}\tablenotemark{c} & 11.7~(+3)~$\pm 2.6$&11.5~(+3)~$\pm
1.5$&10.0~(+1.9)~$\pm 0.8$\\
\scriptsize{\citet{Mada00}}& $9.7^{\tiny{+3.0}}_{\tiny{-1.9}}$ & $9.0^{\tiny{+2.6}}_{\tiny{-1.7}}$ & $7.9^{\tiny{+2.0}}_{\tiny{-1.2}}$  \\
\scriptsize{\citet{Tota01}}& $10.9~\pm~1.1$ & N/A & $8.3~\pm~0.8$ \\
\scriptsize{\citet{Thom03}\tablenotemark{d}}& N/A & 7.0 & N/A \\
\enddata
\tablenotetext{a}{\citet{Mada00} $J-$band is centered at $1.1 \mu$m while this and other studies listed are centered at $1.25~\mu$m.}
\tablenotetext{b}{This work is done at $K_s (2.12~\mu$m) while \citet{Mada00} and \citet{Tota01} are at $K (2.2~\mu$m).}
\tablenotetext{c}{Missing flux estimates are given in parentheses next to our
  measured values for the IGL.  The missing flux in the $J$ and $H-$bands are
  taken from \citet{Thom07a} and for $K_s-$band the value is derived as
  described in Section~\ref{missingflux}.  The confidence intervals given are
  the $1~\sigma$ error estimates from the galaxy counts fits described in Section~\ref{nir_resolved}.}
\tablenotetext{d}{$\rm{Lower~limit}$.}
\end{deluxetable*}

\begin{figure*}
\begin{center}
\includegraphics[width=150mm]{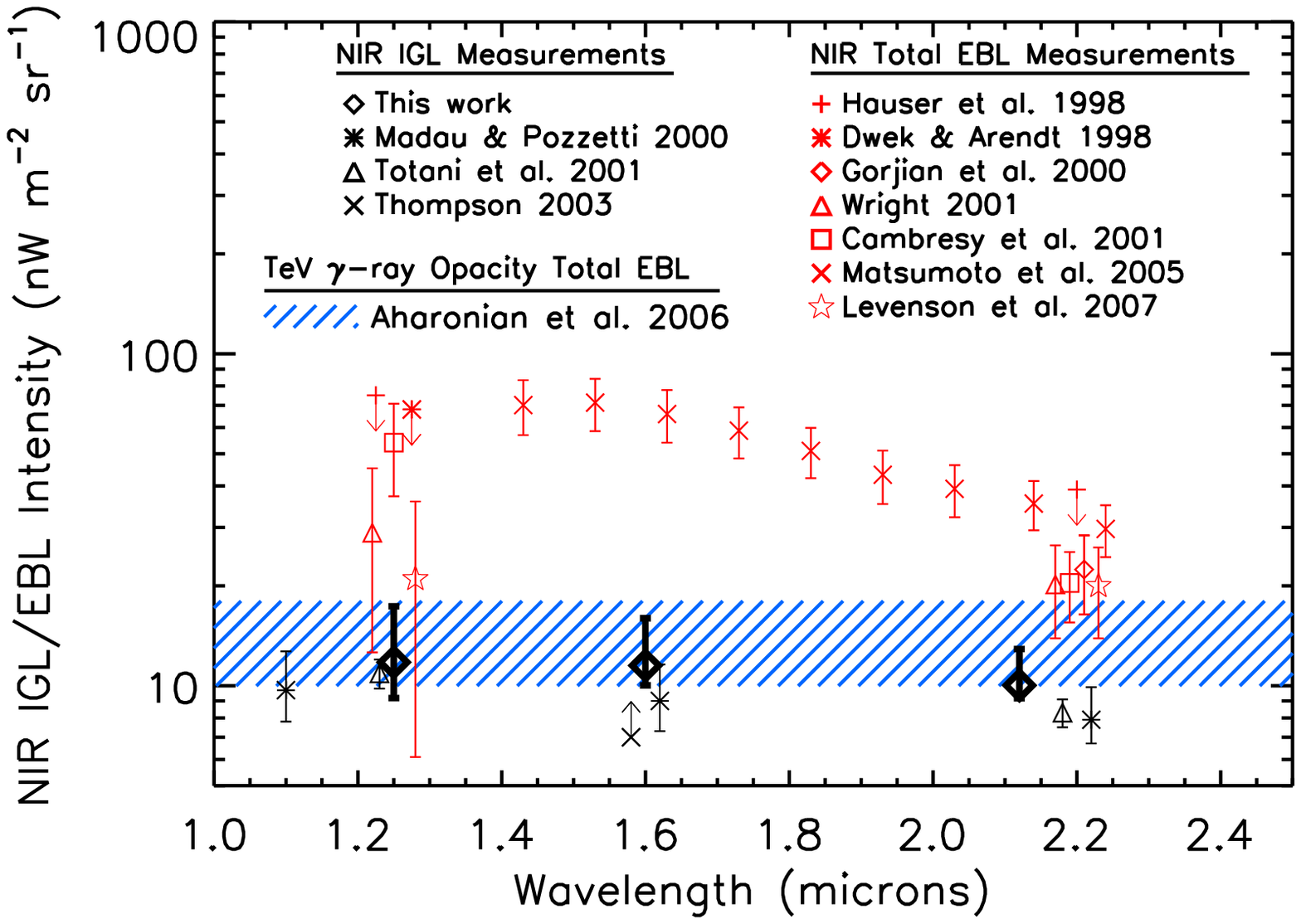}
\caption{\label{all_nirebl}NIR IGL and EBL measurements over the past decade
  as a function of wavelength ($\mu$m).  This work (black diamonds) and a
  summary of measurements of the NIR IGL (other black symbols) via
  integrated galaxy counts \citep{Mada00,Tota01, Thom03}, total EBL (red
  symbols) via total light minus stars and zodiacal light \citep{Dwek98,
    Haus98, Gorj00, Wrig01, Camb01, Mats05, Leve07}, and via TeV $\gamma-$ray
  opacity measurements (blue hashed area) \citep{Ahar06}.  The range indicated
  for the $\gamma-$ray work shows the $1~\sigma$ error range.  Note that data
  points at $1.25, 1.6,$~and~$2.2~\mu$m have been shifted slightly in their
  abscissa values for clarity.  Arrows are used to denote upper and lower
  limits and otherwise error bars represent $1~\sigma$ confidence levels.
  Our results bring the measurement of the IGL into better
  agreement with TeV $\gamma-$ray observations and the most recent total
  total NIR EBL measurements.  The lower limits on our data points show the
  $1~\sigma$ error estimates associated with the galaxy counts integration
  described in Section~\ref{nir_resolved}, while
  the upper limits show these same $1~\sigma$ error estimates plus the missing
  flux component derived in Section~\ref{missingflux} and shown in Table~\ref{nirebl_tab}}
\end{center}
\end{figure*} 
In Figure~\ref{all_nirebl} we show our IGL results alongside a summary of
measurements of the NIR IGL (black symbols) and EBL (red symbols) over the past decade, including results
obtained via integrated galaxy counts \citep{Mada00,Tota01, Thom03}, via
total light minus stars and zodiacal light \citep{Dwek98, Haus98, Gorj00,
  Wrig01, Camb01, Mats05, Leve07}, and via  TeV $\gamma-$ray opacity (blue
hashed area) \citep{Ahar06}.  

The blue hashed area of Figure~\ref{all_nirebl} shows the allowed ($1~\sigma$) NIR EBL
intensity ($14 \pm~4$ nW m$^{-2}$ sr$^{-1}$) derived from High Energy Stereoscopic System (HESS) observations of
TeV blazars \citep{Ahar06}.   \citet{Mazi07} use 13 TeV blazars and a grid of
NIR background intensities to further constrain the NIR EBL and find
approximate agreement with the results of \citet{Ahar06}.  As noted earlier,
however, estimates of the NIR EBL from TeV $\gamma-$ray opacity measurements
rely on assumptions about the intrinsic SEDs of blazars and that of the EBL,
both of which are poorly constrained observationally.  Our results (black diamonds) are some $10-20$\% higher than
previous estimates of the IGL, which puts them closer to EBL estimates from
TeV blazar observations and the most recent total NIR EBL measurements.   

A large NIRBE has been found by several groups, with perhaps
the most striking result being that of \citet{Mats05}.  This excess, when
combined with other NIR and optical background measurements, showed an apparent
spectral break around 1~$\mu$m.  The excess was originally attributed to
PopIII stars, with the break corresponding to the redshifted Lyman limit for
these stars at $z\sim 10$.  However, a search for the possible absorption
imprint of this break on the $\gamma-$ray SED of blazars did not find evidence
for such a feature \citep{Dwek05a}.

\citet{Mada05} discuss the implications of a NIRBE due to
massive PopIII stars in the early universe and demonstrate that an
excess of even a few nW m$^{-2}$ sr$^{-1}$ in the $J-$band would imply
energetic requirements that are ``uncomfortably high''.  They note that $\sim
5\%$ of baryons would be formed into such stars by a redshift of 9 (roughly twice the fraction converted into stars since then) and that the 
metals created in these PopIII stars would have to be hidden in
intermediate mass black holes (IMBHs) for the universe to avoid exceeding
solar metallicity by redshift of 9.  They point out that the IMBHs themselves
could produce the NIR excess as miniquasars, as suggested by \citet{Sant02},
but that accretion onto these IMBHs would dominate the soft X-ray background
and their mass density in the present day universe would exceed that of black
holes found in galactic nuclei by 3 orders of magnitude.  Finally, they show
that the ionizing flux produced by these PopIII stars would exceed by 3
orders of magnitude that required to produce the $WMAP$ observed electron
scattering depth at $z=17$.

\section{Summary}
\label{summary}

We have presented measurements of the IGL in $J,H,$ and $K_s$
using deep wide-field NIR photometry in combination with ultradeep
MOIRCS $K_s-$band data and selections from the literature.  These results
place the best current constraints on the total NIR light from resolved
galaxies and serve as a new lower limit to the total NIR EBL.   While these results are in relative agreement with previous
measurements, our numbers are $10-20$\% higher, bringing them into better
agreement with those derived from $\gamma-$ray experiments and the most recent
measurements of the total NIR EBL.  

We find the IGL to be roughly isotropic, consistent with the expectation of large-scale isotropy in the EBL. We confirm that the starlight
subtraction for the most recent total NIR EBL measurements is correct, so if there still exists a foreground subtraction issue in these
measurements, it most likely is associated with the zodiacal light.  

While our measurements cannot rule out the existence of a NIRBE due to PopIII
stars or other exotic early universe objects, our new lower
limits on the IGL and the upper limits
found from TeV $\gamma-$ray experiments \citep{Ahar06} could now be considered
in rough agreement with the most recent total NIR EBL
measurements in the $J-$band, and in near agreement in the $K-$band.

\acknowledgements{We thank the anonymous referee for their careful review of
  this article and their comments and suggestions that helped to improve the
  content and presentation of this work.  

We gratefully acknowledge support from the NSF grants
AST 0708793 (A.~J.~B.) and AST 0709356
(L.~L.~C.), the University of Wisconsin Research Committee with funds granted
by the Wisconsin Alumni Research Foundation, and the David and Lucile Packard
Foundation (A.~J.~B.).  R.~C.~K. was supported by a
Wisconsin Space Grant Consortium Graduate Fellowship, a Sigma Xi Grant in
Aid of Research and an NSF East Asia and Pacific Summer Institutes Fellowship during portions of this work. 

This publication makes use of data products from the
Two Micron All Sky Survey (2MASS), which is a joint project of the University
of Massachusetts and the Infrared Processing and Analysis Center/California
Institute of Technology, funded by the National Aeronautics and Space
Administration and the National Science Foundation.  

This work is based in
part on data products from the UKIRT Infrared Deep Sky Survey (UKIDSS).

This work is based in part on observations obtained with WIRCam, a joint
project of CFHT, Taiwan, Korea, Canada, France, and the Canada-France-Hawaii
Telescope (CFHT) which is operated by the National Research Council (NRC) of
Canada, the Institute National des Sciences de l'Univers of the Centre
National de la Recherche Scientifique of France, and the University of
Hawaii.  

This work is based in part on data collected at the Subaru
Telescope, which is operated by the National Astronomical Observatory of Japan.

This work has made use of NASA's Astrophysics Data System.

\bibliography{my}
\end{document}